\documentclass[journal]{IEEEtran}
\usepackage{textcomp}
\usepackage{amsfonts}
\usepackage{amssymb}
\usepackage{amsmath}
\usepackage{algorithm}
\usepackage{multirow}
\usepackage{graphicx}
\usepackage{cite}
\usepackage{exscale}
\usepackage{relsize}
\usepackage{algpseudocode}
\usepackage{graphics}
\usepackage{mathrsfs}
\usepackage{siunitx}
\usepackage{url}
\usepackage{lipsum}
\usepackage{multirow}
\usepackage{array}
\usepackage{fancyhdr}

\ifCLASSOPTIONcompsoc
\usepackage[caption=false,font=normalsize,labelfont=sf,textfont=sf]{subfig}
\else
\usepackage[caption=false,font=footnotesize]{subfig}
\fi

\newcommand{\bm}[1]{\mbox{\boldmath{$#1$}}}

\newtheorem{Def}{Definition}
\newtheorem{Theo}{Theorem}
\newtheorem{Prop}{Proposition}

\usepackage[T1]{fontenc}

\def\BibTeX{{\rm B\kern-.05em{\sc i\kern-.025em b}\kern-.08em
    T\kern-.1667em\lower.7ex\hbox{E}\kern-.125emX}}

\begin{document}
\title{Optimal Sampling for Dynamic Complex Networks with Graph-Bandlimited Initialization}

\author{\IEEEauthorblockN{Zhuangkun~Wei,
                          Bin~Li, and Weisi~Guo}

\thanks{Zhuangkun~Wei, and Weisi Guo are with the School of Engineering, the University of Warwick,
West Midlands, CV47AL, UK. (Email: zhuangkun.wei@warwick.ac.uk).}
\thanks{Bin Li is with the School of Information and Communication Engineering (SICE),
Beijing University of Posts and Telecommunications (BUPT), Beijing,
100876, China.}
}


\maketitle

\begin{abstract}
Many engineering, social, and biological complex systems consist of dynamical elements connected via a large-scale network. Monitoring the network's dynamics is essential for a variety of maintenance and scientific purposes. Whilst we understand how to optimally sample a single dynamic element or a non-dynamic graph, we do not possess a theory on how to optimally sample networked dynamical elements. Here, we study nonlinear dynamic graph signals on a fixed complex network. We define the necessary conditions for optimal sampling in the combining time- and graph-domain to fully recover the networked dynamics. We firstly interpret the networked dynamics into a linearized matrix. Then, we prove that the dynamic signals can be sampled and fully recovered if the networked dynamics is stable and their initialization is bandlimited in the graph spectral domain. This new theory directly maps optimal sampling locations and rates to the graph properties and governing nonlinear dynamics. This can inform the optimal placement of experimental probes and sensors on dynamical networks as well as inform the design of each sensor's optimal sampling rate. We motivate the reader with two examples of recovering the networked dynamics for: social population growth and networked protein biochemical interactions with both bandlimited and arbitrary initialization.
\end{abstract}


\thispagestyle{fancy}
\fancyhead{}
\lhead{}
\lfoot{}
\cfoot{\footnotesize{\copyright~2019 IEEE. Personal use of this material is permitted. Permission from IEEE must be obtained for all other uses, in any current or future media, including reprinting/republishing this material for advertising or promotional purposes, creating new collective works, for resale or redistribution to servers or lists, or reuse of any copyrighted component of this work in other works.}}
\rfoot{},

\section{Introduction}
In networked ecosystems, each element has a functional behaviour (e.g. a self-dynamic describable by a differential equation). When individual elements are coupled together via a complex network (with coupling dynamics), the whole networked system can exhibit the necessary complex behaviour \cite{Gao16}. There are many examples of complex networks with explicit (e.g. from phase synchronization \cite{Krishnagopal17} to nonlinear dynamics \cite{Barzel13}) or latent dynamics, spanning: urban structure \cite{Wilson08}, social networks \cite{8105893}, economics \cite{Bardoscia17}, infrastructure \cite{Schafer18}, ecology \cite{Lu16}, biology clocks \cite{Hasegawa18PhyE}, epidemic spreading \cite{Scholtes14}, and organizational structure \cite{Ellinas17}. Whilst many such systems can be described by explicit differential equations (e.g. the mean behaviour is predictable), individual systems will differ due to usage, deterioration, and other factors. This is reflected in parameter uncertainty. As such, data monitoring of the network \cite{Duarte12, Hens18} is essential for both scientific study and maintaining operational capacity. We propose to use the deterministic complex system framework to identify the optimal sampling strategy. In the context of the networked dynamical systems (see Fig.~\ref{fig1}), we do not yet know how to optimally sample the dynamic complex networks from a joint graph- and time-domain perspective. Lack of sampling knowledge on dynamic graphs can lead to over-sampling (expensive) or under-sampling (cannot recover overall behaviour).

\subsection{Literature Review}
Optimal sampling for a single dynamic process is determined by the Nyquist sampling theorem. Optimal sampling on combinatorial graphs is given by the theory of spectral analysis \cite{chung1996spectral}, whereby a specific operator (e.g., the Laplacian operator, \cite{chung1996spectral,p2008}, and the weighted adjacent matrix \cite{Sandryhaila14}) is employed to analyze the spectrum components. Based on these foundations, GSP research in recent years integrates them to understand how to sample graph signals \cite{p2008,Sandryhaila14,anis2014towards,Chen15,wang2015generalized,anis2016efficient,Chen16,wang2018optimal,chamon2018greedy,ortega2018graph}. For instance, \cite{p2008} first introduced the notion of the Paley-Wiener spaces with respect to the operator on combinatorial graphs, and analyzed the graph spectrum of the signals that belong to that space. Further developments proposed the concept of the uniqueness set of nodes that can be used to sample and perfectly recover graph signals. Based on these advances, \cite{anis2014towards,Chen15,anis2016efficient,Chen16, wang2018optimal,chamon2018greedy} provided several methods to find the uniqueness set. More recent work \cite{8115204} considers Joint Fourier Transform (JFT) for dynamic graphs, but the sampling set changes with time, which is not useful for some real-world sensing applications (e.g. optimal sensor deployment).

\begin{figure}[!t]
\centering
\includegraphics[width=3.5in]{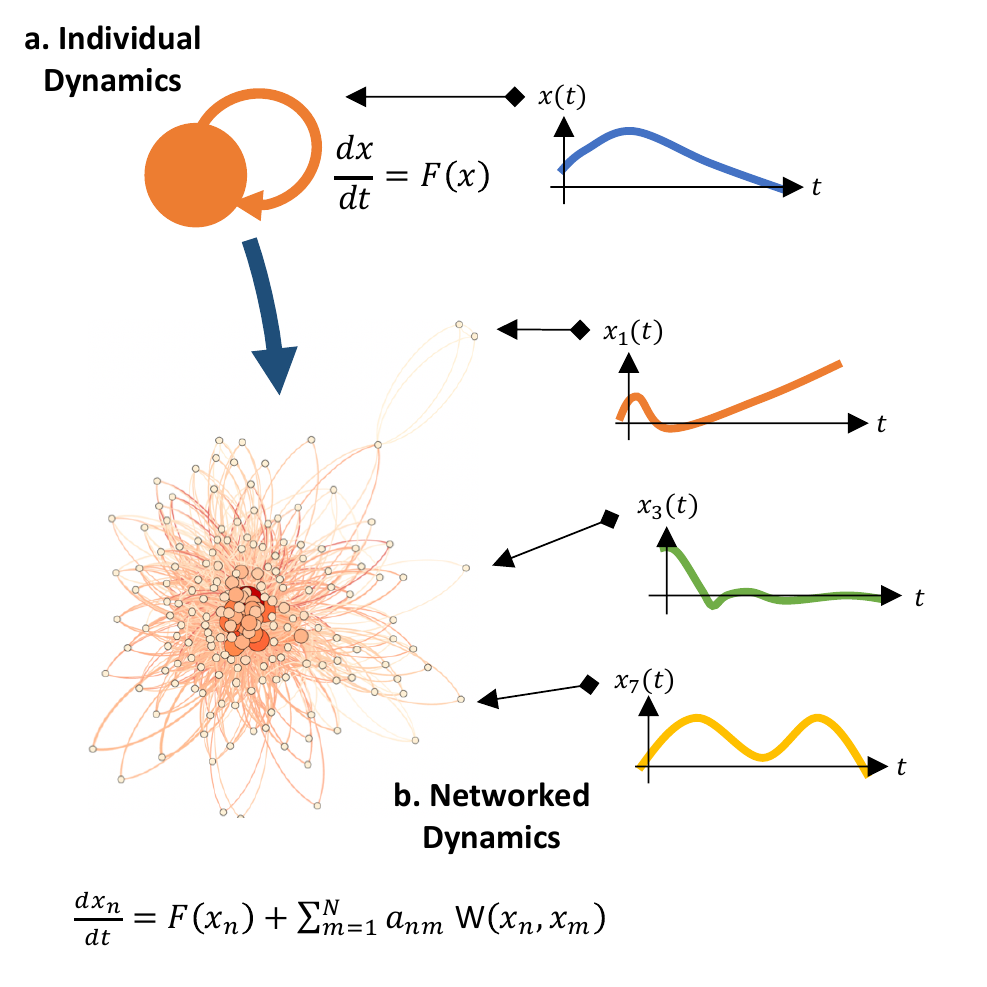}
\caption{Illustration of complex network governed by dynamic equations: (a) individual dynamics, and (b) networked dynamics.}
\label{fig1}
\end{figure}

Whilst these studies contribute to the advancement on how to select a fixed set of sampling nodes, they cannot be used for sampling dynamic graph signals governed by explicit nonlinear dynamics - see Fig.~\ref{fig1}. This is mainly because as the time-varying graph signal evolves, it is difficult to design an operator that is capable of ensuring that all the dynamic graph signals belong to the Paley-Wiener space. Consequently, there remains a lack of understanding on the sample theory of the joint graph- and time-domain.
Different from the compressed sensing (e.g. a tensor) \cite{Sidropoulos12, Ding17}, dynamic graphs on the one hand give explicit knowledge on its structural form, and on the other hand are governed by nonlinear dynamics that have causal relations between their time states. As such, whilst the notion of GSP with explicit dynamics shares similarities with compressed sensing, it differs in its framework and application. Notably in our work, we directly map optimal sampling locations and rates to the graph properties and governing nonlinear dynamics, which doesn't rely on data properties (e.g., the sparse structure) required by the tensor compressed sensing.

\subsection{Contributions \& Organization}
In this work, we suggest a novel sampling theory from the joint time- and graph-domains for dynamic graph signals governed by explicit nonlinear dynamics. The main contributions of this paper are listed as follows.

(1) We linearize the nonlinear networked dynamics, which provides a pathway to finding the optimal graph sample set as well as the cut-off frequency from the time-domain, provided the overall network dynamics satisfy Lyapunov stability \footnote{This is sensible for most stable real-world systems.}. In this view, sampling the dynamic network can be viewed as time sampling on critical nodes.

(2) We provide the theory on computing the graph cut-off frequency and its corresponding sampling node set. With the help of the linearized dynamics from (1), we prove that the linearized dynamic graph signals have the same graph bandwidth with the input. This indicates that we can use the bandwidth as the graph cut-off frequency for loss-less sampling and recovery.

(3) We prove that the graph bandwidth maps to the cut-off frequency from the time-frequency domain. As such, the sampling rate from time-domain can be computed via the Nyquist theorem. More importantly, this relation indicates an \textbf{explicit mapping} of the optimal sampling locations and rates to the graph properties and the governing dynamics. This framework provides the dynamical system insight unavailable from previous GSP and data-driven compressed sensing research. Our fixed optimal sampling nodes also improves over current research \cite{8115204} which yields dynamic sampling nodes.

(4) We evaluate our proposed sampling theory via two different application domains: (a) networked social population with linear dynamics, and (b) protein networks with nonlinear biochemical interactions. We consider both bandlimited and arbitrary inputs and the simulation results demonstrate the successful recovery of the overall networked dynamics with minimal loss. This suggests that the proposed sampling framework is beneficial to a wide range of scientific and engineering applications.

The rest of this paper is structured as follows. In Section II, we detail the networked nonlinear dynamical system model considered. In Section III, we formulate a theory on joint time- and graph-domain optimal sampling and explain the necessary conditions for recovery of the dynamics. In Section IV, we motivate the reader with two examples in networked social population dynamics and networked protein biochemical interactions. In Section V, we conclude the paper and discuss potential future areas of research. \\


\section{System Model}
In this section, we provide a concept of dynamic graph signals governed by dynamic equations.

Signal processing on dynamic graphs is concerned with the analysis and processing of signal-set where dynamic signal elements are connected to each other with respect to both the graph topology and the nonlinear dynamics on the node and edge (as is shown in Fig.~\ref{fig1}). The relation is expressed through the graph $G(\mathcal{V}, \mathbf{W})$, in which $\mathcal{V}=\{v_1,\cdots,v_{N}\},N\in\mathbb{N}^+$ represents a set of nodes, and $\mathbf{W}$ is the adjacency matrix of the graph $G(\mathcal{V}, \mathbf{W})$. For each node $v_n\in\mathcal{V}$, $\alpha_{n,m}\in\{0,1\}$ of a directed edge from $v_m$ to $v_n$ reflects the connectivity from the $m$th signal element to the $n$th one. Therefore, the time-varying signal $x_n(t)$ corresponding to $v_n$ can be described as evolving in accordance to its self-dynamic function $f_n(\cdot)$ and the mutualistic coupling function $g_{n,m}(x_n(t),x_m(t))$ - see Fig.~\ref{fig1}:
\begin{equation}
\label{dynamic equation}
\frac{dx_n(t)}{dt}=f_n(x_n(t))+\sum_{m=1}^N\alpha_{n,m}\cdot g_{n,m}(x_n(t),x_m(t)).
\end{equation}
With the help of Eq. (\ref{dynamic equation}), we here denote the graph signal set as $\mathbf{x}(t)=[x_1(t),\cdots,x_N(t)]^T$. Unlike the traditional graph signal that considers only a fixed data on the graph \cite{p2008, Sandryhaila14, anis2014towards, Chen15, wang2015generalized, anis2016efficient, Chen16, wang2018optimal, chamon2018greedy, ortega2018graph}, we extend the concept to the time-varying signals governed by the dynamical equations.

\begin{figure*}[!t]
\centering
\includegraphics[width=7in]{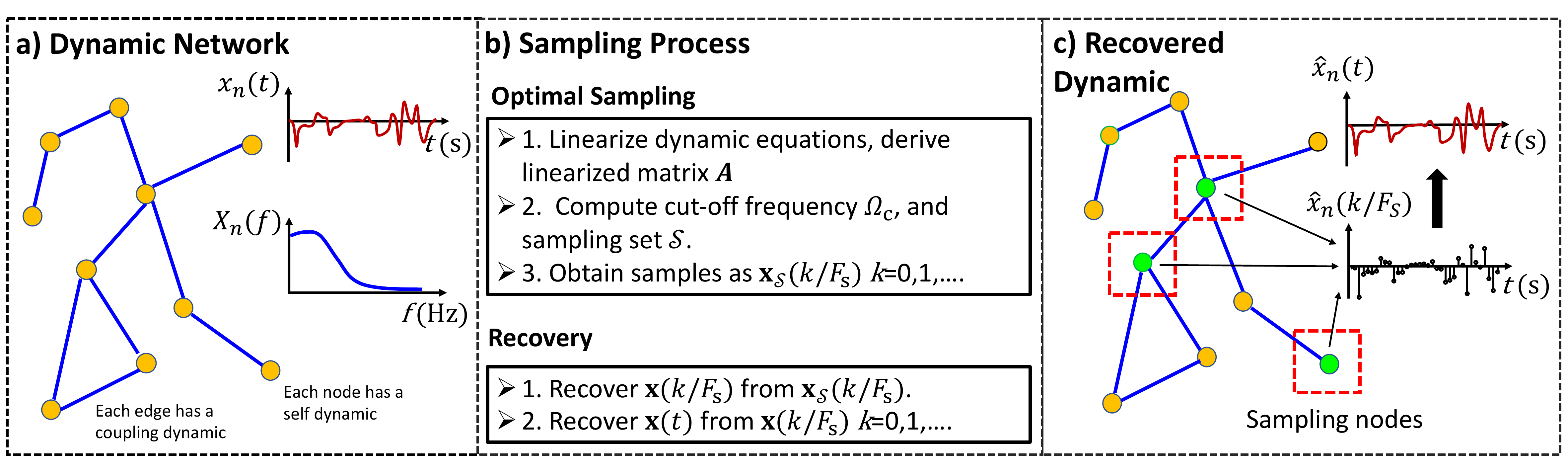}
\caption{Illustration of: (a) complex network with dynamic graph signals from governing nonlinear dynamics; (b) the sampling and recovering process in terms of the combination of the time- and the graph-domain; and (c) the recovered network dynamics.}
\label{fig2}
\end{figure*}

Whilst many such systems given by Eq. (\ref{dynamic equation}) can be described by explicit differential equations (e.g. the mean behaviour is predictable), individual systems will differ due to usage, deterioration, and other factors. This is reflected in parameter uncertainty. As such, data monitoring of the network \cite{Duarte12, Hens18} is essential for both scientific study and maintaining operational capacity. Therefore, the purpose of this paper is to study how to sample and recover the signals in the dynamic network from the combining time and graph -domains. To be specific, we should compute the sample frequency, denoted as $F_\text{s}$ via the time-frequency perspective, as well as consider which nodes should be regarded as the sampling nodes (shown in Fig.~\ref{fig2}), i.e., the composition of the sampling node set, denoted as $\mathcal{S}\subset\mathcal{V}$. Also, we should ensure a fixed sampling node set $\mathcal{S}$ that does \textbf{not} vary with time, as changing the sensor deployment in some real applications (e.g., the under-water surveillance) may be impractical.


\section{Sampling for Dynamic Graph Signal}
In this section, we elaborate our sampling theory on dynamic graph signals. In order to ensure an existence of time-frequency Fourier transform of $x_n(t)$, we introduce the limitation of Lyapunov stability on our networked system, which is reasonable for most real systems. As such, this enables us to convert the nonlinear dynamics to its linear approximation.

\subsection{Sketch of Sampling on Dynamic Graph Signals}


As we assume an existence of the equilibrium point $\mathbf{x}_e$ of Eq. (\ref{dynamic equation}), we introduce the definition of Lyapunov stability as follows.
\begin{Def}
\cite{lyapunov1992general} We say $\mathbf{x}(t)$ has the asymptotical Lyapunov stability on point $\mathbf{x}_e$, if and only if for any $\epsilon>0$, there exists a $\delta>0$ such that, if $||\mathbf{x}(0)-\mathbf{x}_e||<\delta$, then for every $t>0$ we have
\begin{equation}
||\mathbf{x}(t)-\mathbf{x}_e||<\epsilon,~\text{and}~ \lim_{t\rightarrow+\infty}||\mathbf{x}(t)-\mathbf{x}_e||=0.
\label{Lyapunov}
\end{equation}
\end{Def}
In this paper, we consider only the dynamic graph signal $\mathbf{x}(t)$ that is asymptotically Lyapunov stable. From Eq. (\ref{Lyapunov}), we know that $\mathbf{x}(t)$ converges to $\mathbf{x}_e$, and therefore $\mathbf{x}(\infty)=\mathbf{x}_\text{e}$. For convenience, we specify:
\begin{equation}
\mathbf{y}(t)=\mathbf{x}(t)-\mathbf{x}(\infty),
\end{equation}
and we have $\lim_{t\rightarrow\infty}\mathbf{y}(t)=\mathbf{y}(\infty)=\mathbf{0}$.

Also, given that $\mathbf{y}(t)$ is Lyapunov stable on $\mathbf{y}(\infty)=\mathbf{0}$, analyzing the group of nonlinear differential functions in Eq. (\ref{dynamic equation}) can be viewed as studying its linear approximations as follows:
\begin{equation}
\frac{dy_n(t)}{dt}=\mathbf{A}\cdot\mathbf{y}(t)+o\left(\|\mathbf{y}(t)\|\right)
\approx\mathbf{A}\cdot\mathbf{y}(t),
\end{equation}
where $o\left(\|\mathbf{y}(t)\|\right)$ are terms that go to zeros faster than the first order for $t\rightarrow+\infty$. $\mathbf{A}$ represents the Jacobian matrix at point $\mathbf{x}(\infty)$, i.e.,
\begin{equation}
\begin{aligned}
\mathbf{A}=\mathbf{W}\circ&\begin{bmatrix}\frac{\partial g_{1,1}(t)}{\partial x_{1}(t)} & \cdots&\frac{\partial g_{1,N}(t)}{\partial x_N(t)}\\ \vdots & \ddots & \vdots\\ \frac{\partial g_{N,1}(t)}{\partial x_1(t)} & \cdots &\frac{\partial g_{N,N}(t)}{\partial x_N(t)}\\
\end{bmatrix}\Bigg|_{\mathbf{x}(t)=\mathbf{x}(\infty)}\\
&+diag\left\{\frac{\partial{F_1}}{\partial{x_1}},\cdots,\frac{\partial{F_N}}{\partial{x_N}}\right\}\Bigg|_{\mathbf{x}(t)=\mathbf{x}(\infty)},
\end{aligned}
\label{linearized}
\end{equation}
where $\circ$ represents the \textit{Hadamard product}.

It is noteworthy that $\mathbf{A}$ has two merits. First, $\mathbf{A}$ combines both the topology of the network (as is represented by the adjacency matrix $\mathbf{W}$), as well as the dynamic equations (i.e., the mutualistic coupling functions and the self-dynamics defined in Eq. (\ref{dynamic equation})). Second, given the condition of Lyapunov stability, $\mathbf{A}$ is capable of approximating the non-linear dynamic networked signal with minor errors. In this view, we transform the non-linear dynamic into its linear approximation, and thus fruitful matrix theory can be relied on to analyze the joint time-graph sampling theory.

Next, we will focus on deriving both the cut-off frequency from the time-domain, denoted as $\Omega_\text{c}$, and the graph sampling set, $\mathcal{S}$.

\subsubsection{Time-frequency Fourier Analysis}
From the perspective of sampling via the time- and frequency-domain, one critical demand is to compute the cut-off frequency $\Omega_{\text{c}}$, so that the Nyquist sampling theory leveraged on $\Omega_{\text{c}}$ can be adopted. To do so, we need to firstly analyze the time-frequency Fourier transform, and the Theorem is as follows.
\begin{Theo}
if $\mathbf{y}(t)$ has the Lyapunov stability on $\mathbf{y}(\infty)=\mathbf{0}$, then $y_n(t)$ is integrable, and its time-frequency Fourier transform exists, i.e.,
\begin{equation}
Y_n(\Omega)=\int_{0}^{+\infty}y_n(t)\cdot e^{-i\Omega t}dt.
\end{equation}
\end{Theo}

The Theorem 1 proves the existence of the time-frequency Fourier transform of $y_n(t)$. Therefore, it provides a premise that the graph signals $\mathbf{y}(t)$ can be sampled and recovered via the time-domain, provided that one can derive the cut-off frequency $\Omega_\text{c}$. Then, by assigning the sampling frequency as $F_\text{s}\geq\Omega_\text{c}/\pi$, the signal $y_n(t)$ can be sample as $y_n(k/F_\text{s})$, and then recovered as:
\begin{equation}
\hat{y}_n(t)=\sum_{k\in\mathbb{Z}}y_n\left(\frac{k}{F_\text{s}}\right)\cdot sinc\left(F_\text{s}\left(t-\frac{k}{F_\text{s}}\right)\right),
\end{equation}
where $sinc(t)=\sin\pi t/(\pi t)$ is the interpolation function.

\subsubsection{Graph Fourier Analysis}
From the graph-domain, we aim to find a sampling set $\mathcal{S}\subset\mathcal{V}$, so the signal $\mathbf{y}(k/F_\text{s})$ in Eq. (7) can be recovered via the samples on nodes in $\mathcal{S}$, i.e., $\mathbf{y}_\mathcal{S}(k/F_\text{s})=[y_\sigma(k/F_\text{s})]^T,\sigma\in\mathcal{S}$.
To do so, we borrow the idea of the graph spectrum analysis \cite{chung1996spectral,p2008}. In essence, for any graph signal $\mathbf{y}(k/F_{\text{s}})$, if there exists an \textit{operator}, denoted as $\bm{\Delta}$, to whom $\mathbf{y}(k/F_{\text{s}})$ is bandlimited, then the sampling set $\mathcal{S}$ can be selected using the operator. In the context of the dynamic graph signal processing, we require the operator $\bm{\Delta}$ to have following properties.

\textbf{(i)} The operator $\bm{\Delta}$ should be able to be diagonalized or Jordan decomposed \cite{Sandryhaila14}, where the consequent eigenvalues are referred as the graph frequencies \cite{ortega2018graph,Sandryhaila14}, and the eigenvectors represent the Graph Fourier bases \cite{ortega2018graph,Sandryhaila14}. This guarantees that, the spectrum of $\mathbf{y}(k/F_\text{s})$ with respect to $\bm{\Delta}$ can be characterized by the graph frequencies with their corresponding bases. As such, the selection of sampling set $\mathcal{S}$ can be pursued by the existing theory in \cite{anis2014towards,Chen15,anis2016efficient,Chen16, wang2018optimal,chamon2018greedy}.

\textbf{(ii)} For all the $k$, the operator $\bm{\Delta}$ should maintain that $\mathbf{y}(k/F_\text{s})$ have the \textbf{same} set of non-zero graph frequency components. In this way, both the sampling set $\mathcal{S}$ and the recovering process will not change with respect to the dynamics, which is reasonable for most of the sampling process in the complex network monitoring applications.

Note that, compared with (i), (ii) is more difficult to satisfy. In this paper, we provide the satisfied $\bm{\Delta}$ only in the case of $\mathbf{y}(k/F_\text{s})$ is bandlimited from the graph-domain. Hence, before we study how to find the suitable $\bm{\Delta}$ for (ii), we firstly give the definition of $\bm{\Delta}$ that follows the (i), and the definition of graph bandlimited signal.

\begin{Def}
$\bm{\Delta}$ is a matrix operator such that $\bm{\Delta}: \mathbb{R}^N\rightarrow\mathbb{R}^N$
\end{Def}
The reasoning behind our matrix operator is that we want $\bm{\Delta}$ to be either diagonalizable or can be Jordan decomposed. Hence, there exists a normalized non-singular matrix $\bm{\Gamma}=[\bm{\gamma}_1,\cdots,\bm{\gamma}_N]$ corresponding to a matrix of eigenvalues (either a diagonal matrix or a Jordan form of quasi-diagonal matrix), and we can use them for graph-frequency analysis. For convenience, we assume $\bm{\Delta}$ is diagonalizable, and the following theories for Jordan decomposition is similar. As such, we have \cite{ortega2018graph}
\begin{equation}
\bm{\Delta}=\bm{\Gamma}\cdot diag\{\lambda_1,\cdots,\lambda_N\}\cdot\bm{\Gamma}^{-1}.
\label{decom}
\end{equation}
In Eq. \eqref{decom}, we denote the eigenvectors $\bm{\gamma}_1,\cdots,\bm{\gamma}_N$ as the graph Fourier bases, with their corresponding eigenvalues $\lambda_1,\cdots,\lambda_N(\lambda_j\in\mathbb{C})$ as the discrete graph frequencies \cite{ortega2018graph,Sandryhaila14}. In this view, the graph signal $\mathbf{y}(k/F_\text{s})$ can be decomposed by the graph frequencies and their corresponding bases, i.e.,
\begin{equation}
    \tilde{\mathbf{y}}\left(\frac{k}{F_\text{s}}\right)=\bm{\Gamma}^{-1}\cdot\mathbf{y}\left(\frac{k}{F_\text{s}}\right),
\end{equation}
where $\tilde{\mathbf{y}}(k/F_\text{s})=[\tilde{y}_1(k/F_\text{s}),\cdots,\tilde{y}_N(k/F_\text{s}]^T$ is referred as the Graph Fourier transform. Hence, similar to the definition of the band-limit from the time-domain, the graph-domain bandlimited signal can be defined as long as we know which graph frequency parts are corresponding to the low and high -frequency. Here, we use the concept of smoothness via the quadratic form of variation, and prove that the frequencies with smaller variation are lower frequencies.

\begin{Theo}\cite{Sandryhaila14}
For $\forall\lambda_j,\lambda_m\in\{\lambda_1,\cdots,\lambda_N\}$, $Var(\bm{\Delta},\bm{\gamma}_j)<Var(\bm{\Delta},\bm{\gamma}_m)$ if and only if $|\lambda_j-|\lambda|_{\text{max}}|<|\lambda_m-|\lambda|_{\text{max}}|$,
with the quadratic form of variation of $\bm{\Delta}$ as:
\begin{equation}
Var\left(\bm{\Delta},\mathbf{y}\left(\frac{k}{F_\text{s}}\right)\right)=\frac{1}{2}\left\|\mathbf{y}\left(\frac{k}{F_\text{s}}\right)-\bm{\Delta}_{\text{norm}}\cdot\mathbf{y}\left(\frac{k}{F_\text{s}}\right)\right\|_2^2,
\end{equation}
where $|\lambda|_\text{max}$ is the maximum magnitude of $\{|\lambda_1|,\cdots,|\lambda_N|\}$, $\|\cdot\|_2$ denotes the 2-norm, and $\bm{\Delta}_{\text{norm}}=\bm{\Delta}/|\lambda|_{\text{max}}$.
\end{Theo}

From Theorem 2, we understand that the frequency bases $\bm{\gamma}_j$ whose $\lambda_j$ is closer to $|\lambda|_\text{max}$ on the complex plane, is smoother (known as the lower graph-domain frequency) with respect to $\bm{\Delta}$. In this view, as we assign a non-negative $\omega$ (referred as the \textbf{bandwidth}), we can determine the graph-domain frequency values that belong to the smooth parts corresponding to $\omega$.
\begin{Def}
Given $\bm{\Delta}$, for a given bandwidth $\omega>0$, we define the graph-domain $\omega$-bandlimited frequency set as
\begin{equation}
\lambda_\omega(\bm{\Delta})=\left\{\lambda_j||\lambda_j-|\lambda|_\text{max}|<\omega,\lambda_j\in\{\lambda_1,\cdots,\lambda_N\}\right\}.
\end{equation}
\end{Def}

From the Definition 3, we map the bandwidth $\omega$ to the graph-domain frequency set $\lambda_\omega(\bm{\Delta})$ (as is shown in Fig. 3(a)). For example, $\lambda_{+\infty}(\bm{\Delta})=\{\lambda_1,\cdots,\lambda_N\}$, and $\lambda_0(\bm{\Delta})=\varnothing$. With the help of Definition 3, we can define the graph-domain bandlimited signal constructed by the bases whose frequencies belongs to the $\omega$-bandlimited frequency set, $\lambda_\omega(\bm{\Delta})$.
\begin{Def}
We say $\mathbf{y}(k/F_\text{s})$ is $\omega$-bandlimited with respect to $\bm{\Delta}$, if and only if:
\begin{equation}
\begin{aligned}
\mathbf{y}\left(\frac{k}{F_\text{s}}\right)&=\sum_{j\in\mathcal{N}_\omega}\tilde{y}_j\left(\frac{k}{F_\text{s}}\right)\cdot\bm{\gamma}_j,\\
&\text{with}~\mathcal{N}_{\omega}=\{j|\lambda_j\in\lambda_\omega(\bm{\Delta})\}.
\end{aligned}
\end{equation}
\end{Def}
As we know the cut-off bandwidth of $\mathbf{y}(k/F_\text{s})$ is $\omega_\text{c}$ corresponding to $\bm{\Delta}$, we can select the sampling set $\mathcal{S}$ via \cite{anis2014towards,Chen15,anis2016efficient,Chen16, wang2018optimal,chamon2018greedy} such that:
\begin{equation}
rank\left(\bm{\Gamma}_{\mathcal{S}\mathcal{N}_{\omega_\text{s}}}\right)=|\mathcal{N}_{\omega_\text{s}}|,
\end{equation}
for any $\omega_\text{s}\geq\omega_\text{c}$, where $\bm{\Gamma}_{\mathcal{S}\mathcal{N}_{\omega_\text{s}}}$ denotes the matrix composed by the elements whose rows belong to $\mathcal{S}$ and whose columns are in $\mathcal{N}_{\omega_\text{s}}$. Hence, the dynamic signal $\mathbf{y}(k/F_\text{s})$ can be sampled as $\mathbf{y}_\mathcal{S}(k/F_\text{s})$, and fully recovered as:
\begin{equation}
\hat{\mathbf{y}}\left(\frac{k}{F_\text{s}}\right)=\bm{\Phi}\cdot\mathbf{y}_{\mathcal{\mathcal{S}}}\left(\frac{k}{F_\text{s}}\right),
\end{equation}
where $\bm{\Phi}=\bm{\Gamma}_{\mathcal{V}\mathcal{N}_{\omega_{\text{s}}}}\cdot\left(\bm{\Gamma}^T_{\mathcal{S}\mathcal{N}_{\omega_{\text{s}}}}\bm{\Gamma}_{\mathcal{S}\mathcal{N}_{\omega_{\text{s}}}} \right)^{-1}\bm{\Gamma}^T_{\mathcal{S}\mathcal{N}_{\omega_{\text{s}}}}$.

\subsubsection{Dynamic Signal Recovery from graph- and time- domains}
Given Eq. (7) and Eq. (15) that consider the signal recovery from the time- and the graph- domains respectively, we here give the combined recovery process (which is illustrated in Fig. 2(c)). We write the sample matrix from both the time- and graph- domain as $\mathbf{Y}_\mathcal{S}=[\mathbf{y}_\mathcal{S}(0/F_\text{s}),\cdots,\mathbf{y}_\mathcal{S}(K/F_\text{s})]$. The interpolation matrix for time-domain is given as $\bm{\Psi}=[sinc(F_\text{s}(t-0/F_\text{s})),\cdots,sinc(F_\text{s}(t-K/F_\text{s}))]^T$. Hence, the recovered signal $\hat{\mathbf{y}}(t)$ can be computed as:
\begin{equation}
\hat{\mathbf{y}}(t)=\bm{\Phi}\cdot\mathbf{Y}_\mathcal{S}\cdot\bm{\Psi}.
\end{equation}

Next, we will study how to determine the time-domain cut-off frequency, i.e., $\Omega_\text{c}$, and the graph-domain cut-off frequency, i.e., $\omega_\text{c}$, along with the way to find the suitable $\bm{\Delta}$ that satisfies (ii). To do so, we analyze the case of graph-domain bandlimited initialization.

\subsection{Graph-domain Bandlimited Initialization}
With the help of the linear approximation of the dynamics in Eqs. (4)-(5), we here consider the case where the initialization $\mathbf{y}(0)$ is $\omega$-bandlimited \textbf{with respect to $\mathbf{A}$}.

\subsubsection{Sampling from Graph Domain}
\begin{Theo}
As $\mathbf{y}(t)$ follows the 1-order and linear differential equation, i.e., $d(\mathbf{y}(t))/dt=\mathbf{A}\mathbf{y}(t)$, and has Lyapunov stability on the equilibrium $\mathbf{0}$, if $\mathbf{y}(0)$ is $\omega$-bandlimited with respect to $\mathbf{A}$, then $\mathbf{y}(t)$ is $\omega$-bandlimited with respect to $\mathbf{A}$
\end{Theo}

\begin{IEEEproof}
The $\mathbf{y}(t)$ can be computed as
\begin{equation}
\mathbf{y}(t)=e^{t\cdot\mathbf{A}}\cdot\mathbf{y}(0).
\end{equation}
And we have
\begin{equation}
\begin{aligned}
\tilde{\mathbf{y}}(t)=&\mathbf{U}^{-1}\cdot\mathbf{y}(t)\\
=&\mathbf{U}^{-1}\cdot\sum_{k=0}^{\infty}\frac{t^k}{k!}\cdot\mathbf{A}^k\cdot\mathbf{y}(0)\\
=&\sum_{k=0}^{+\infty}\frac{t^k}{k!}\cdot diag\{\mu_1,\cdots,\mu_N\}^k\cdot\mathbf{U}^{-1}\cdot\mathbf{y}(0)\\
=&\sum_{k=0}^{+\infty}\frac{t^k}{k!}\cdot diag\{\mu_1^k,\cdots,\mu_N^k\}\cdot\tilde{\mathbf{y}}(0),
\end{aligned}
\end{equation}
with $\mathbf{A}=\mathbf{U}\cdot diag\{\mu_1,\cdots,\mu_N\}\cdot\mathbf{U}^{-1}$, and $\mathbf{A}^0=\mathbf{I}$. This suggests that the subscripts of the non-zero elements of $\tilde{\mathbf{y}}(t)$ coincide with those of $\tilde{\mathbf{y}}(0)$. Therefore, it proves that $\mathbf{y}(t)$ is $\omega$-bandlimited with respect to $\mathbf{A}$..

\end{IEEEproof}
\begin{Prop}
As $d(\mathbf{y}(t))/dt=\mathbf{A}\mathbf{y}(t)$, and $\mathbf{y}(t)$ has Lyapunov stability on the equilibrium $\mathbf{0}$, if $\mathbf{y}(0)$ is $\omega$-bandlimited with respect to $\mathbf{A}$, then the operator $\bm{\Delta}$ can be determined as $\bm{\Delta}=\mathbf{A}$.
\end{Prop}
\begin{Prop}
As $d(\mathbf{y}(t))/dt=\mathbf{A}\mathbf{y}(t)$, and $\mathbf{y}(t)$ has Lyapunov stability on the equilibrium $\mathbf{0}$, if $\mathbf{y}(0)$ is $\omega$-bandlimited with respect to $\mathbf{A}$, then the graph-domain cut-off frequency $\omega_\text{c}$ for signal $\mathbf{y}(t)$ can be selected as:
\begin{equation}
    \max\{|\lambda_{\omega}(\mathbf{A})-|\lambda|_\text{max}|\}\leq\omega_{\text{c}}\leq\omega.
\end{equation}
\end{Prop}
\begin{IEEEproof}
According to Theorem 2, $\max\{|\lambda_{\omega}(\mathbf{A})-|\lambda|_\text{max}|\}$ represents the largest distance from $|\lambda|_\text{max}$ for all eigenvalues that belongs to the $\omega$-bandlimited frequency set with respect to $\mathbf{A}$, i.e.,  $\lambda_{\omega}(\mathbf{A})$. Hence, we have $\lambda_\omega(\mathbf{A})\equiv\lambda_{\omega_\text{c}}(\mathbf{A})$ for any $\omega_\text{c}$ satisfying Eq. (18). This is also illustrated via Fig. 3(a), in which the graph-domain cut-off frequency can be selected from the area such that the graph-domain frequency parts with non-zero coefficients (i.e., $\lambda_1$, $\lambda_4$, and $\lambda_5$) will not change
\end{IEEEproof}

From the Theorem 3 and its two Propositions, we can derive the suitable matrix operator as $\bm{\Delta}=\mathbf{A}$, and the graph-domain cut-off frequency $\omega_\text{c}$. In this view, the sampling set $\mathcal{S}$ can be selected according to Eq. (14), and hence we can obtain the samples from nodes in $\mathcal{S}$ as $\mathbf{y}_{\mathcal{S}}(k/F_\text{s})$, and further recover $\mathbf{y}(k/F_\text{s})$ from Eq. (15).

\subsubsection{Sampling from Time Domain}
After we compute the graph-domain cut-off frequency $\omega_\text{c}$, we will deduce the time-domain cut-off frequency $\Omega_{\text{c}}$.
\begin{Theo}
As $d(\mathbf{y}(t))/dt=\mathbf{A}\mathbf{y}(t)$, and $\mathbf{y}(t)$ has Lyapunov stability on the equilibrium $\mathbf{0}$, if $\mathbf{y}(0)$ is $\omega$-bandlimited with respect to $\mathbf{A}$, then the time-frequency Fourier transform of $y_n(t)$ is:
\begin{equation}
Y_n(\Omega)=\sum_{\lambda_j\in\lambda_\omega(\mathbf{A})}\frac{\gamma_{n,j}\cdot\tilde{y}_j(0)}{-\text{Re}[\lambda_j]+i\left(\Omega-\text{Im}[\lambda_j]\right)}.
\end{equation}
\end{Theo}
\begin{IEEEproof}
Without losing the generality, we order the eigenvalues with the descending of $Var(\bm{\Delta},\gamma_m)$. Given from the Proposition 1 that $\bm{\Delta}=\mathbf{A}$, we hereby compute Eq. (16) as:
\begin{equation}
\begin{aligned}
&\mathbf{y}(t)\\
=&\mathbf{\Gamma}\cdot diag\{e^{\lambda_1t},\cdots,e^{\lambda_Nt}\}\cdot\mathbf{\Gamma}^{-1}\cdot\mathbf{y}(0)\\
=&\mathbf{\Gamma}\cdot diag\{e^{\lambda_1t},\cdots,e^{\lambda_Nt}\}\cdot
[0,\cdots,0,~\underset{\min{j},s.t.\lambda_{j}\in\lambda_\omega(\mathbf{A})}{\underbrace{\tilde{y}_{j}(0),\cdots,\tilde{y}_{N}(0)}}]^T\\
=&\mathbf{\Gamma}\cdot[0,\cdots,0,~\underset{\min{j},s.t.\lambda_{j}\in\lambda_\omega(\mathbf{A})}{\underbrace{\tilde{y}_{j}(0)\cdot e^{\lambda_{j}t},\cdots,\tilde{y}_{N}(0)\cdot e^{\lambda_{N}t}}}]^T,
\end{aligned}
\end{equation}
which suggests that each element of $\mathbf{y}(t)$ can be described as a summation of weighted $e^{\lambda_{j}t},\cdots,e^{\lambda_Nt}$. Therefore, we have $y_n(t)$ as follows:
\begin{equation}
y_n(t)=\sum_{\lambda_j\in\lambda_\omega(\mathbf{A})}\gamma_{n,j}\cdot\tilde{y}_j(0)\cdot e^{\lambda_jt}.
\end{equation}
Given that $\mathbf{y}(t)$ is Lyapunov stable, the eigenvalues of $\bm{\Delta}=\mathbf{A}$ have non-positive real values \cite{lyapunov1992general}, i.e., $\text{Re}[\lambda_j]\leq0$. Hence, its time-frequency Fourier transform is computed as:
\begin{equation}
\begin{aligned}
Y_n(\Omega)=&\int_{0}^{+\infty}\sum_{\lambda_j\in\lambda_\omega(\mathbf{A})}\gamma_{n,j}\cdot\tilde{y}_j(0)\cdot e^{\lambda_jt}\cdot e^{-i\Omega t}dt\\
=&\sum_{\lambda_j\in\lambda_\omega(\mathbf{A})}\gamma_{n,j}\cdot\tilde{y}_j(0)\int_{0}^{+\infty}e^{\text{Re}[\lambda_j]t-i\left(\Omega-\text{Im}[\lambda_j]\right) t}dt\\
=&\sum_{\lambda_j\in\lambda_\omega(\mathbf{A})}\frac{\gamma_{n,j}\cdot\tilde{y}_j(0)}{-\text{Re}[\lambda_j]+i\left(\Omega-\text{Im}[\lambda_j]\right)}.
\end{aligned}
\end{equation}
\end{IEEEproof}

With the help of Theorem 4, we can observe from Eq. (19) that the time-frequency Fourier transform relates to the eigenvalues of $\mathbf{A}$, which are also the graph-domain frequencies. This indicates that we can rely on the computation of the graph-domain cut-off frequency $\omega_{\text{c}}$, to deduce the time-domain cut-off frequency $\Omega_c$ by taking $\omega_\text{c}$ into Eq. (19).

\begin{figure*}[!t]
\centering
\includegraphics[width=6in]{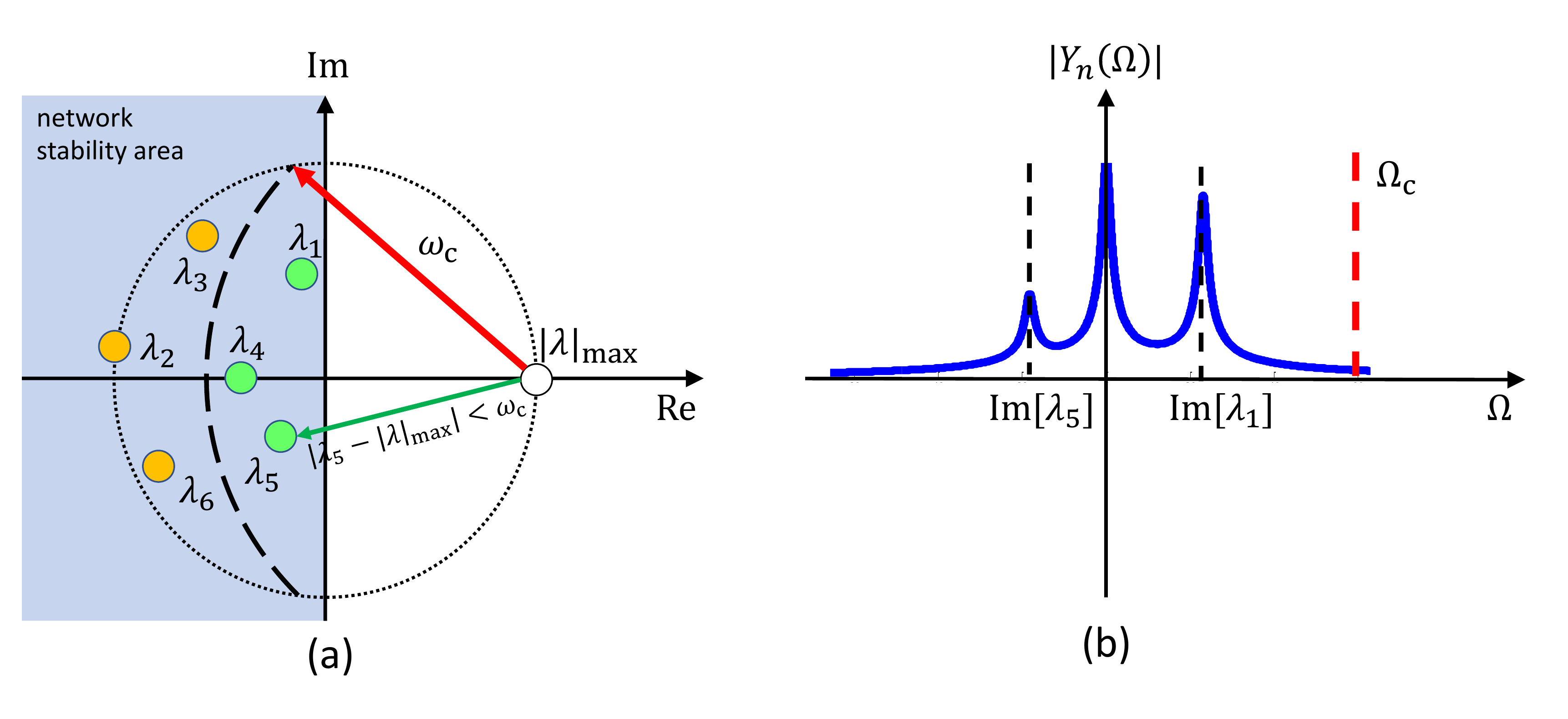}
\caption{Illustration of the relations between graph-domain cut-off frequency $\omega_\text{c}$, the time-domain cut-off frequency $\Omega_\text{c}$, and the underlying graph structure and its governing nonlinear dynamics $\mathbf{A}$. Subplots: (a) network stability area for the eigenvalues of $\mathbf{A}$, along with the graph-domain frequencies that belong to the $\omega$-bandlimited frequency set $\lambda_\omega(\bm{\Delta})$ (i.e., the Definition 3); and (b) the magnitude of time-frequency Fourier transform $|Y_n(\Omega)|$ for the graph-domain $\omega_\text{c}$-bandlimited signals. It can be seen that the eigenvalues of $\mathbf{A}$ inside the graph-domain $\omega$-bandwidth (i.e., $\lambda_1,\lambda_4,\lambda_5$) determines the shape of $|Y_n(\Omega)|$. This further demonstrates that the time-domain cut-off frequency $\Omega_\text{c}$ can be determined by Proposition 3.}
\label{figK}
\end{figure*}

\begin{Prop}
Consider $d(\mathbf{y}(t))/dt=\mathbf{A}\mathbf{y}(t)$, $\mathbf{y}(t)$ has Lyapunov stability on the equilibrium $\mathbf{0}$, and $\mathbf{y}(0)$ is $\omega$-bandlimited with respect to $\mathbf{A}$. Given that we are only interested in frequency components of time-domain larger than a threshold $\varepsilon$, as long as the graph-domain cut-off frequency $\omega_\text{c}$ is known, the time-domain cut-off frequency $\Omega_\text{c}$ can be computed as:
\begin{equation}
\begin{aligned}
\Omega_{\text{c}}=\max&\{\left|\text{Im}[\lambda_{\omega_\text{c}}(\mathbf{A})]\right|\}\\
&+\sqrt{\frac{\|\mathbf{y}(0)\|_2^2}{\varepsilon^2}-\min\{\text{Re}[\lambda_{\omega_\text{c}}(\mathbf{A})]\}^2}
\end{aligned}
\label{Final}
\end{equation}
where $\max\{\left|\text{Im}[\lambda_{\omega_\text{c}}(\mathbf{A})]\right|\}$ represents the maximum absolute of the imaginary part of graph-domain frequencies in $\lambda_{\omega_\text{c}}(\mathbf{A})$, and $\min\{\text{Re}[\lambda_{\omega_\text{c}}(\mathbf{A})]\}$ is the minimum real parts of frequencies in $\lambda_{\omega_\text{c}}(\mathbf{A})$.
\end{Prop}

\begin{IEEEproof}
The magnitude of $Y_n(\Omega)$ is computed as:
\begin{equation}
\begin{aligned}
\left|Y_n(\Omega)\right|=&\left|\sum_{\lambda_j\in\lambda_\omega(\mathbf{A})}\frac{\gamma_{n,j}\cdot\tilde{y}_j(0)}{-\text{Re}[\lambda_j]+i\left(\Omega-\text{Im}[\lambda_j]\right)}\right|\\
\leq&\sum_{\lambda_j\in\lambda_\omega(\mathbf{A})}\left|\frac{\gamma_{n,j}\cdot\tilde{y}_j(0)}{-\text{Re}[\lambda_j]+i\left(\Omega-\text{Im}[\lambda_j]\right)}\right|\\
\leq&\sum_{\lambda_j\in\lambda_\omega(\mathbf{A})}\frac{|\gamma_{n,j}\cdot\tilde{y}_j(0)|}{\sqrt{\text{Re}^2[\lambda_j]+\left(\Omega-\text{Im}(\lambda_j)\right)^2}}.
\end{aligned}
\end{equation}
We can learn from Eq. (24) that the imaginary parts of the eigenvalues belong to $\lambda_\omega(\mathbf{A})$ contributes to the left/right shift of $\Omega$. Hence, for any $\Omega>\max\{\left|\text{Im}[\lambda_{\omega_\text{c}}(\mathbf{A})]\right|\}$, we have:
\begin{equation}
\begin{aligned}
&\left|Y_n(\Omega)\right|\\
<&\frac{\|\mathbf{y}(0)\|_2}{\sqrt{\min\{\text{Re}[\lambda_{\omega_\text{c}}(\mathbf{A})]\}^2+\left(\Omega-\max\{\left|\text{Im}[\lambda_{\omega_\text{c}}(\mathbf{A})]\right|\}\right)^2}}.
\end{aligned}
\end{equation}
Then, by making the right-hand of Eq. (25) smaller than $\varepsilon$, we can deduce the Eq. (23).
\end{IEEEproof}

\begin{Prop}
As $d(\mathbf{y}(t))/dt=\mathbf{A}\mathbf{y}(t)$, and $\mathbf{y}(t)$ has Lyapunov stability on the equilibrium $\mathbf{0}$, if $\mathbf{y}(0)$ is $\omega$-bandlimited with respect to $\mathbf{A}$, given the threshold $\varepsilon>0$, and the graph-domain cut-off frequency $\omega_\text{c}$, the time-domain cut-off frequency $\Omega_\text{c}$ has an upper-bound related to $\omega_\text{c}$:
\begin{equation}
\Omega_{\text{c}}<\min\left\{\sqrt{\omega_\text{c}^2-|\lambda|_\text{max}^2},\frac{\omega_\text{c}\sqrt{4|\lambda|_\text{max}^2-\omega_\text{c}^2}}{2|\lambda|_\text{max}}\right\}+\frac{\|\mathbf{y}(0)\|_2}{\varepsilon}.
\end{equation}
\end{Prop}

\begin{IEEEproof}
This can be proved as we replace $\max\{\left|\text{Im}[\lambda_{\omega_\text{c}}(\mathbf{A})]\right|\}$ with the minimum $Im$ of the intersection between $Re^2+Im^2=|\lambda|_\text{max}^2$, $(Re-|\lambda|_\text{max})^2+Im^2=\omega_\text{c}^2$, and $Re=0$, as well as replacing $\min\{\text{Re}[\lambda_{\omega_\text{c}}(\mathbf{A})]\}^2$ with $0$ (as is shown in Fig. 3(a)).
\end{IEEEproof}

\subsubsection{Explicit Relationship between Optimal Sampling and Graph Dynamics}
It is important to stress that a key benefit of our framework is the creation of an \textbf{explicit relationship} between the time- and graph-domain cut-off frequencies, the graph properties, and the nonlinear dynamics. We will elaborate this relationship from three aspects.

(i) The networked dynamics characterized by the self-dynamic function and the mutualistic coupling equations in Eq. (1) is interpreted by the linearized matrix $\mathbf{A}$ from Eq. (5). For example, the stability of the network can be analyzed via the real parts of the eigenvalues of $\mathbf{A}$ (seen from Fig.~\ref{figK}(a)).

(ii) In the case of the graph-domain $\omega$-bandlimited initialization with respect to $\mathbf{A}$, the graph-domain cut-off frequency of the dynamic signals i.e., $\omega_\text{c}$ can be computed via $\omega$ as Eq. (18). This $\omega_\text{c}$ maps the eigenvalue (graph-domain frequency) set $\lambda_{\omega_\text{c}}(\mathbf{A})$ (as is illustrated in Fig.~\ref{figK}(a)), such that only the bases (eigenvectors) whose eigenvalues belongs to $\lambda_{\omega_\text{c}}(\mathbf{A})$ have non-zero coefficients.

(iii) This graph-domain cut-off $\omega_\text{c}$ further leads to the computation of the time-domain cut-off frequency, as only the eigenvalues of $\mathbf{A}$ that belongs to $\lambda_{\omega_\text{c}}(\mathbf{A})$ affect the shape of the time-frequency Fourier transform (seen from Fig.~\ref{figK}(b)). Also, a direct relation of $\Omega_\text{c}$ and $\omega_\text{c}$ is shown via Eq. (26).
In summery, referring to Eq. (18), Eqs. (23)-(26), and illustrated in Fig.~\ref{figK}, we can see that the time-domain cut-off frequency $\Omega_{\text{c}}$ is related to the eigenvalues that belong to $\lambda_{\omega_\text{c}}(\mathbf{A})$, which in turn is related to the optimally sampled graph structure and the underlying dynamics.

\subsection{General Case with Arbitrary Initialization}
It is noteworthy that in real applications on dynamic complex networks, $\mathbf{y}(0)$ is $\omega$-bandlimited with respect to $\mathbf{A}$ may not be easily satisfied. Therefore, monitoring the network with arbitrary input $\mathbf{y}(0)$ is still challenging, as the operator $\bm{\Delta}$ is difficult to design, which makes the computation of reliable $\omega_\text{c}$ and $\Omega_\text{c}$ challenging. Here, inspired by the sampling theories deduced with the bandlimited signals, we provide the imperfect sampling method for this case from both the time and graph -domain.

\subsubsection{Sampling from Graph Domain}
From the perspective of graph sampling, an intuitive idea is to regard $\mathbf{y}(0)$ as an approximated $\omega$-bandlimited signal. From Eq. (17), we learn that the graph frequency parts of $\tilde{\mathbf{y}}(t)$ varies independently with each other, i.e.,
\begin{equation}
\tilde{y}_n(t)=\sum_{k=0}^{+\infty}\frac{t^k}{k!}\cdot \mu_n^k\cdot\tilde{y}_n(0).
\end{equation}
Hence, we can approximate $\mathbf{y}(0)$ via selecting $|\mathcal{S}|\leq N$ maximum amplitudes from $\tilde{\mathbf{y}}(0)$.

\begin{Theo}
As $d(\mathbf{y}(t))/dt=\mathbf{A}\mathbf{y}(t)$, and $\lim_{t\rightarrow\infty}\mathbf{y}(t)=\mathbf{0}$, the operator $\bm{\Delta}=\mathbf{A}$ can be applied for graph sampling and recovery with an error as:
\begin{equation}
\|\hat{\mathbf{y}}(t)-\mathbf{y}(t)\|_2\leq\|\mathbf{y}(0)\|_2\cdot\frac{1-|\mathcal{S}|}{N}.
\end{equation}
\end{Theo}

\begin{figure*}[!t]
\centering
\includegraphics[width=7in]{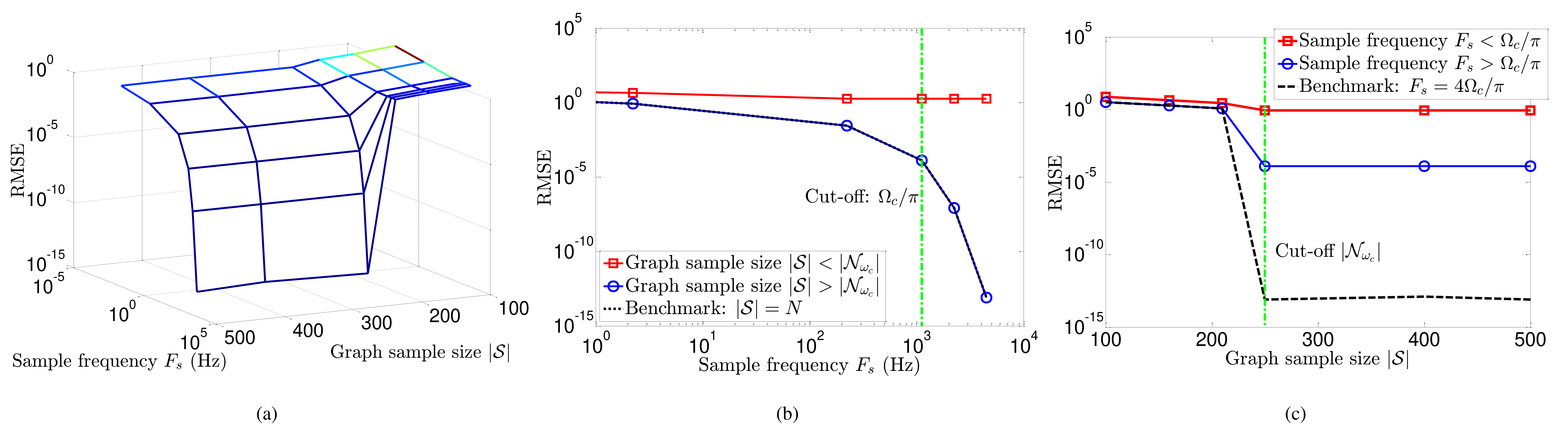}
\caption{Performance of the sampling theory on PD \textbf{linear dynamic network} with \textbf{bandlimited initialization}: (a) RMSE with respect to both the sample frequency $F_s$ and the graph sample size $|\mathcal{S}|$, (b) illustrates one tangent plane of (a), RMSE related to the sample frequency $F_s$ with fixed graph sample size $|\mathcal{S}|$, and (c) shows another tangent plane, the RMSE affected by only the graph sample size $|\mathcal{S}|$.}
\label{fig3}
\end{figure*}

\subsubsection{Sampling from Time Domain}
Given that as $\mathbf{y}(0)$ is not $\omega$-bandlimited with respect to $\mathbf{A}$, the sampling theory from Theorem 4 and Proposition 3 cannot hold, as Eq. (20) can no longer describe $y_n(t)$. In this case, the expression of $y_n(t)$ can be written as:
\begin{equation}
y_n(t)=\sum_{\lambda_j\in\lambda_{+\infty}(\mathbf{A})} \gamma_{n,j}\cdot\tilde{y}_j(0)\cdot e^{\lambda_jt}.
\end{equation}
And we can write the time-frequency Fourier transform as:
\begin{equation}
Y_n(\Omega)=\sum_{\lambda_j\in\lambda_{+\infty}(\mathbf{A})} \frac{\lambda_{n,j}\cdot\tilde{y}_j(0)}{-\text{Re}[\lambda_j]+i\left(\Omega-\text{Im}[\lambda_j]\right)}.
\end{equation}
Hence, a similar sampling theory from the time-domain is obtained by changing the Theorem 4 as follows.
\begin{Theo}
For any arbitrary $\mathbf{y}(0)$, given the sampling threshold as $\varepsilon$, the $\Omega_\text{c}$ can be computed as:
\begin{equation}
\begin{aligned}
\Omega_{\text{c}}=\max&\{\left|\text{Im}[\lambda_{+\infty}(\mathbf{A})]\right|\}\\
&+\sqrt{\frac{\|\mathbf{y}(0)\|_2^2}{\varepsilon^2}-\min\{\text{Re}[\lambda_{+\infty}(\mathbf{A})]\}^2}
\end{aligned}
\end{equation}
\end{Theo}
As such, by computing the graph-domain cut-off frequency $\omega_\text{c}$, and time-domain cut-off frequency $\Omega_\text{c}$, the sampling and the recovering processes can be achieved. We now motivate the reader with some example optimal sampling results of real networked systems. \\


\section{Simulations}

In the following analysis, the performance of our proposed sample theory will be evaluated. First, we examine the performance in the case when $\mathbf{y}(0)$ is bandlimited with respect to $\mathbf{A}$. Then, the general cases in which $\mathbf{y}(0)$ is not bandlimited is evaluated. To do so, we specify the root mean square error (RMSE) of $\hat{\mathbf{x}}(t),t\in[0,T]$, i.e.,
\begin{equation}
\begin{aligned}
\text{RMSE}&=\mathbb{E}\{\hat{\mathbf{x}}(t)-\mathbf{x}(t)\}\\
&\backsimeq \sqrt{\frac{\Delta_t}{NT}\sum_{k=0}^{T/\Delta_t-1}\|\hat{\mathbf{x}}(k\Delta_t)-\mathbf{x}(k\Delta_t)\|_2^2},
\end{aligned}
\end{equation}
where $\Delta_t$ is the sample rate whose corresponding time-domain frequency is much greater than the cut-off frequency, i.e., $1/\Delta_t=4\Omega_c/\pi$.

We also use the phrase \textbf{cut-off graph sample size} corresponding to the graph-domain cut-off frequency $\omega_\text{c}$ as $|\mathcal{N}_{\omega_\text{c}}|$, such that $rank\left(\bm{\Gamma}_{\mathcal{S}\mathcal{N}_{\omega_\text{c}}}\right)=|\mathcal{N}_{\omega_\text{c}}|$.

The involved dynamic functions in this simulation are configured as follows. As far as both the linear and nonlinear dynamic networks are concerned, we consider two types of dynamic models \cite{Barzel13}:
\begin{equation}
\label{PD}
\frac{dx_n(t)}{dt}=-Bx_n(t)+R\sum_{m=1}^N \alpha_{n,m}\cdot x_m(t),
\end{equation}
\begin{equation}
\label{MAK}
\frac{dx_n(t)}{dt}=F-Bx_n(t)+R\sum_{m=1}^N\alpha_{n,m}\cdot x_n(t)\cdot x_m(t).
\end{equation}
Eq. (\ref{PD}) is referred as the linear population dynamics (PD) model, where each node's population has a self growth rate $-B$ and also depends on the migration strength $R$ from neighbouring connected nodes. Eq. (\ref{MAK}) is referred as the MAK model that describes the nonlinear dynamics of protein-protein interactions captured by mass-action kinetics. The detailed parameters and explanations for the differential equations are found in \cite{Barzel13}. In Eqs. (\ref{PD})-(\ref{MAK}), we assign the number of nodes $N=500$, and leave other parameters randomly configured such that they satisfy the Lypunov stability defined in Definition 1.

\subsection{Performance with Bandlimited Initialization}
We first evaluate the performance for the graph-domain bandlimited initialization by studying the RMSE in both the linear model and the nonlinear model.

\subsubsection{Linear model}
We illustrate the performance of the sampling theory via the combining time- and graph- domains from Fig.~\ref{fig3}. The Fig.~\ref{fig3}(a) provides the RMSE with respect to the joint time-domain sample frequency $F_\text{s}$ and the graph sample size $|\mathcal{S}|$. We can see that with the increases of both $F_\text{s}$ and $|\mathcal{S}|$, the RMSE reduces, suggesting that the performance of recovering the dynamic graph signals improves, as more samples from both the time- and the graph- domains are involved. This can be also demonstrated in Fig.~\ref{fig3}(b) and Fig.~\ref{fig3}(c), whereby the two tangent planes of Fig.~\ref{fig3}(a) are given.

Fig.~\ref{fig3}(b) investigates the changes of the RMSEs that varied by the time-domain sample frequency $F_s$, as the various fixed graph sample sizes $|\mathcal{S}|$ are considered. With the growth of $F_\text{s}$, the RMSEs are decreasing. Also, it compares the cases whether the fixed graph sample size is larger than the cut-off graph sample size, i.e., $|\mathcal{S}|\gtrless|\mathcal{N}_{\omega_\text{c}}|$. In the case of $|\mathcal{S}|>|\mathcal{N}_{\omega_\text{c}}|$, the RMSE decreases from $10^{-1}$ to $10^{-15}$ as $F_\text{s}$ grows from $10^0$ to $10^{-4}$. This is because that, an increase of time-domain sample frequency $F_\text{s}$ means a growth number of samples from the time-domain, which improves the performance of recovery, according to the Nyquist sample theory. It is also noteworthy that the RMSE from a graph sample size no lesser than the cut-off graph sample size i.e., $|\mathcal{S}|\geq|\mathcal{N}_{\omega_\text{c}}|$ equals to the Benchmark whereby all the nodes are sampled i.e., $|\mathcal{S}|=N$. This is because that in a linear model e.g., the PD model, a perfect recovery can be achieved as $rank\left(\bm{\Gamma}_{\mathcal{S}\mathcal{N}_{\omega_\text{c}}}\right)=|\mathcal{N}_{\omega_\text{c}}|$ is reaching, which validates the Theorem 3 and its Propositions. Then, we consider the case where $|\mathcal{S}|<|\mathcal{N}_{\omega_\text{c}}|$. We can observe that the RMSE reduces little with the growth of the time-domain sample frequency $F_\text{s}$, as the perfect recovery cannot be realized if graph sample size is smaller than the cut-off sample size.

Fig.~\ref{fig3}(c) illustrates the RMSEs with respect to the graph sample size $|\mathcal{S}|$ with the fixed time-domain sample frequencies. As aforementioned. we can firstly observe that the RMSEs decrease with the growth of $|\mathcal{S}|$. Then, we can see that the RMSE from the case $F_\text{s}>\Omega_\text{c}/\pi$ is lower as opposed to that from $F_\text{s}<\Omega_\text{c}/\pi$, since the recovery performance of the latter case are deteriorated by the lack of samples from time-domain. Also, the threshold $\varepsilon$ of the magnitude of the frequency transform from Proposition 3 matters, as the computation process of the cut-off frequency $\Omega_{\text{c}}$ neglects the frequency parts whose magnitudes are smaller than the threshold. This leads to the gap between the benchmark with a larger $F_s=4\Omega_\text{c}/\pi$ and the RMSE whose $F_s=\Omega_\text{c}/\pi$. Furthermore, we can notice that, after the graph sample size reaches the cut-off sample size i.e., $|\mathcal{S}|=|\mathcal{N}_{\omega_c}|$ the RMSEs converges to a limitation. The reason can be categorized as that, in the case of a linear model such as the PD model, Theorem 3 holds true. In other words, if the initialization $\mathbf{y}(0)$ is $\omega_\text{c}$-bandlimited with respect to $\mathbf{A}$, then $\mathbf{y}(t)$ are $\omega_\text{c}$-bandlimited with respect to $\mathbf{A}$, which suggests that we can use any $\mathcal{S}$ as a sampling set such that $rank\left(\bm{\Gamma}_{\mathcal{S}\mathcal{N}_{\omega_\text{s}}}\right)=|\mathcal{N}_{\omega_\text{s}}|\geq|\mathcal{N}_{\omega_\text{c}}|$, and the signals can be perfectly recovered.

\subsubsection{Nonlinear model}

\begin{figure*}[!t]
\centering
\includegraphics[width=7in]{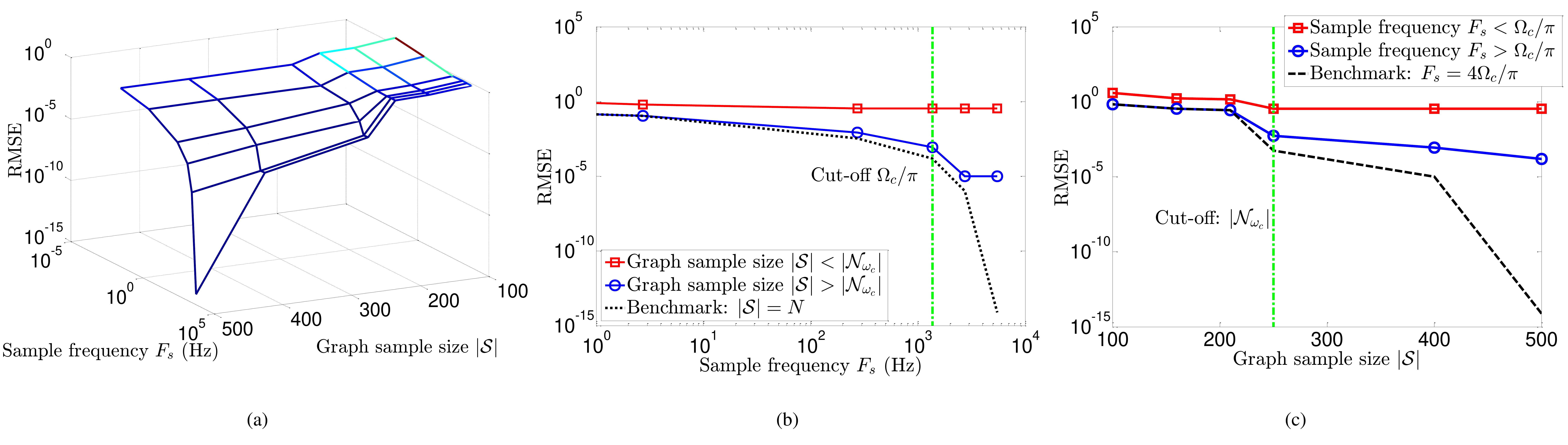}
\caption{Performance of the sampling theory on MAK \textbf{nonlinear dynamic network} with \textbf{bandlimited initialization}: (a) RMSE with respect to both the sample frequency $F_s$ and the graph sample size $|\mathcal{S}|$, (b) illustrates the tangent plane of (a), RMSE corresponding to the sample frequency $F_s$ as the graph sample size $|\mathcal{S}|$ is fixed, and (c) gives another tangent plane, the RMSE influenced by only the effect of graph sample size $|\mathcal{S}|$.}
\label{fig4}
\end{figure*}

The performance of the sampling theory on via the joint time- and graph- domains is illustrated from Fig.~\ref{fig4}. The Fig.~\ref{fig4}(a) shows the RMSE with respect to both the time-domain sample frequency $F_\text{s}$ and the graph sample size $|\mathcal{S}|$. It is intuitive that as $F_\text{s}$ and $|\mathcal{S}|$ increase, the RMSE keeps decreasing, which suggests that the recovery of dynamic graph signals become better as more samples are considered. This can be also demonstrated in Fig.~\ref{fig4}(b) and Fig.~\ref{fig4}(c), whereby the two tangent planes of Fig.~\ref{fig4}(a) are provided.

Fig.~\ref{fig4}(b) illustrates RMSEs that are influenced only by the time-domain sample frequency $F_\text{s}$, as we fix the graph sample size $|\mathcal{S}|$. We can observe that with the increase of $F_\text{s}$, the RMSEs are reduced. Moreover, it compares the cases whether the fixed graph sample size is larger than the cut-off graph sample size, i.e., $|\mathcal{S}|\gtrless|\mathcal{N}_{\omega_\text{c}}|$. For the case $|\mathcal{S}|>|\mathcal{N}_{\omega_\text{c}}|$, the RMSE decreases at first (from $10^{-1}$ to $10^{-5}$), and then converges to a limit (as $10^{-5}$), which is different from Fig.~\ref{fig3}(b). The reasons are given as follows. Firstly, as the time-domain sample frequency $F_\text{s}$ becomes larger, the performance of recovery improves given by the Nyquist sample theory. Then, given the linear approximation of the nonlinear network in Eq. (5), there exists an error limitation, which cannot be surpassed only by increasing $F_s$. This limitation (computed as $10^{-5}-10^{-15}$) is illustrated by the benchmark in Fig.~\ref{fig4}(b) that uses samples of all the nodes from the graph i.e., $|\mathcal{S}|=N$. It is also noteworthy that after $F_\text{s}$ surpasses the cut-off frequency $\Omega_c/\pi$, the RMSE are still decreasing. This is because the computation of $\Omega_\text{c}$ from Proposition 2 ignore the frequency parts whose amplitudes are smaller that the threshold $\varepsilon$. Also, we consider the case where $|\mathcal{S}|<|\mathcal{N}_{\omega_\text{c}}|$. We can observe that the RMSE reduces little as the sample frequency grows. This is because if graph sample size is small, we cannot recover the signals for all $N$ nodes, which limits the RMSE from being decreased.

Fig.~\ref{fig4}(c) gives the RMSEs with respect to the graph sample size $|\mathcal{S}|$ via the fixed time-domain sample frequencies. As aforementioned, the RMSEs decrease with the growth of $|\mathcal{S}|$. Then, we can see that the RMSE from the case $F_\text{s}>\Omega_\text{c}/\pi$ outperforms the one from $F_\text{s}<\Omega_\text{c}/\pi$, as the error from the latter case are restricted by the lack of samples from the time-domain. Also, the threshold of the magnitude of the frequency transform from Proposition 3 matters, as the computed cut-off frequency $\Omega_{\text{c}}$ ignores those whose amplitudes are lesser than the threshold. This gives rise to the gap between the benchmark with a larger $F_\text{s}=4\Omega_\text{c}/\pi$ and the RMSE whose $F_\text{s}=\Omega_\text{c}/\pi$. In addition, we can notice that unlike the Fig.~\ref{fig4}(c) where the RMSEs converges to a limitation after the graph sample size is greater than the cut-off sample size i.e., $|\mathcal{S}|\geq|\mathcal{N}_{\omega_c}|$, the RMSEs are still reducing. The reason can be categorized as that, the graph sample theory we deduced in Theorem 3 is based on the linear system. It is true that the Theorem is still suitable for the nonlinear cases if their dynamic functions satisfy the Lyapunov stability, but the high-order term in Eq. (5) yields the error unless all the nodes from the graph are sampled, i.e., $|\mathcal{S}|=N$. Also, this error gap can be demonstrated and computed via the Benchmark whose sample size is $|\mathcal{S}|=N$.

\subsection{Performance with Arbitrary Initialization}
\begin{figure*}[!t]
\centering
\includegraphics[width=7in]{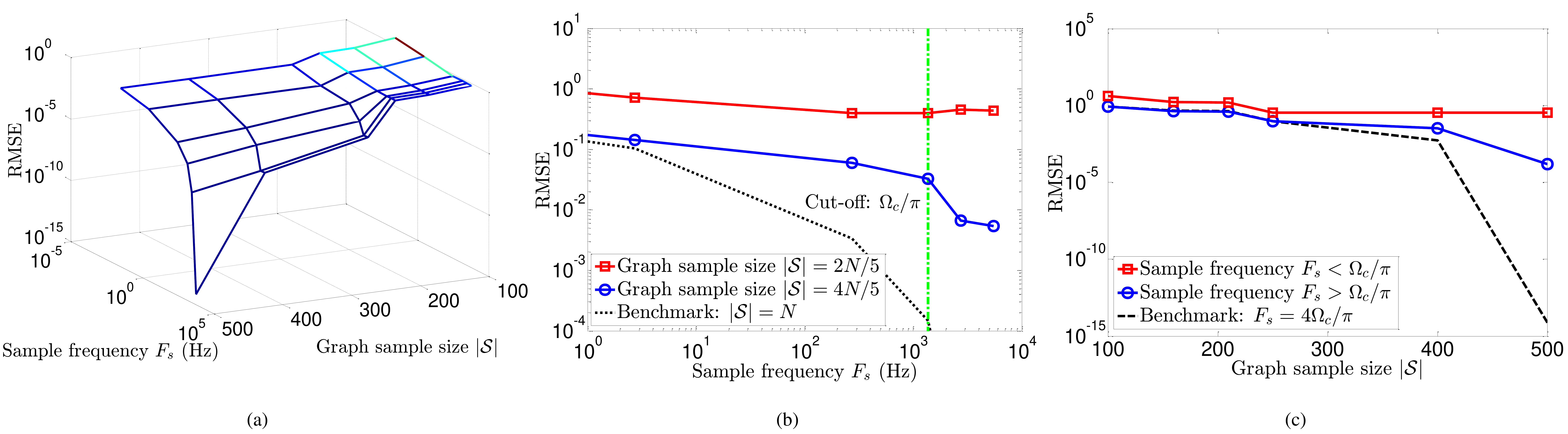}
\caption{Performance of the sampling theory on MAK \textbf{nonlinear dynamic network} with \textbf{arbitrary initialization}: (a) RMSE with respect to both the sample frequency $F_s$ and the graph sample size $|\mathcal{S}|$, (b) illustrates one tangent plane of (a), RMSE related to the sample frequency $F_s$ with fixed graph sample size $|\mathcal{S}|$, and (c) shows another tangent plane, the RMSE affected by only the graph sample size $|\mathcal{S}|$.}
\label{fig5}
\end{figure*}

We then examine our proposed sampling method for arbitrary initialization via the nonlinear model of Eq. (35). The performance of signal recovery is shown in Fig.~\ref{fig5}. The Fig.~\ref{fig5}(a) provides the RMSE varying with both the time-domain sample frequency $F_\text{s}$ and the graph sample size $|\mathcal{S}|$. Similar to the bandlimited cases, the RMSE decreases as $F_\text{s}$ and $|\mathcal{S}|$ increase, demonstrating a higher successful recovery of dynamic graph can be achieved, as more samples are applied.

Fig.~\ref{fig5}(b) illustrates RMSEs with respect to only the time-domain sample frequency $F_\text{s}$, as fixed the graph sample sizes are considered. It is observed that with the increase of $F_\text{s}$, the RMSEs are reducing. Then, it gives the result of comparison between different graph sample sizes, i.e., $|\mathcal{S}|=4N/5$ and $|\mathcal{S}|=2N/5$. We can easily see that the graph sample size $|\mathcal{S}|$ serves as the basic condition for the performance of recovery. This is because if graph sample size is small (i.e., $|\mathcal{S}|=2N/5$), we cannot recover the signals for all $N$ nodes, rendering the RMSE as a constant. Furthermore, we analyze the case with larger graph sample size, i.e.,  $|\mathcal{S}|=4N/5$. We can observe that the RMSE decreases at first (from $10^{-1}$ to $10^{-2}$), and then converges to a limit (as $10^{-2}$), which is higher than that from Fig.~\ref{fig4}(b). The reasons can be categorized as follows. Firstly, just like the bandlimited cases, the performance of recovery improves with a larger time-domain sample frequency $F_\text{s}$, and the aforementioned limitation of the linear approximation contributes to its convergence. Second, different from the bandlimited cases, given that we select $|\mathcal{S}|$ maximum magnitudes from the graph frequency signal $\tilde{\mathbf{y}}(0)$, an error is inevitable as we neglect parts of the dynamics when sampling and recovering the signals. This limitation (computed as $10^{-2}-10^{-15}$) is illustrated by the benchmark in Fig.~\ref{fig5}(b) that uses samples from all the nodes from the graph i.e., $|\mathcal{S}|=N$. We should also note that after $F_\text{s}$ surpasses the cut-off frequency $\Omega_\text{c}/\pi$, the RMSE are still decreasing. This is because the computation of $\Omega_\text{c}$ from Theorem 6 ignore the frequency components whose magnitudes are smaller that the threshold $\varepsilon$.

Fig.~\ref{fig5}(c) illustrates the RMSEs with respect to the graph sample size $|\mathcal{S}|$ where the time-domain sample frequencies $F_\text{s}$ are fixed. As aforementioned, the RMSEs decrease with the growth of $|\mathcal{S}|$. Yet the reason is not only the linear approximation as the bandlimited cases have, the approximation of arbitrary initialization to the bandlimited one in Theorem 5 matters, as we ignore the minimum dynamics from the graph frequency domain.
Then, we can see that the RMSE from the case $F_\text{s}>\Omega_\text{c}/\pi$ outperforms the that from $F_\text{s}<\Omega_\text{c}/\pi$, since the error from the latter case are restricted by the lack of samples from time-domain. Also, the gap between the benchmark with a larger $F_\text{s}=4\Omega_\text{c}/\pi$ and the RMSE whose $F_\text{s}=\Omega_\text{c}/\pi$ is obvious. This is mainly because the computation of the cut-off frequency $\Omega_{\text{c}}$ ignores the parts whose magnitudes are lesser than the threshold.  \\


\section{Conclusions \& Future Research}
In this paper, we developed a theory for the optimal time- and graph-domain joint sampling of a networked dynamical system with bandlimited initialization in the graph spectral domain. We first interpret the networked dynamics into a linearized matrix, from which the stability of the network can be analyzed via its eigenvalues. Then, We prove that the dynamic signals can be sampled and fully recovered if the network are stable and their initialization are bandlimited with respect to the matrix.

Unlike other high-dimensional data sets, we consider dynamical graphs with nonlinear dynamics that have explicit causal relations between nodes. Therefore, our sampling theory is able to directly map optimal graph sampling locations and rates to the graph properties and governing nonlinear dynamics. Changes in the underlying dynamics or the network structure will be able to directly inform the optimal data sampling process (see Fig.~\ref{figK}). Together with recent advances in understanding how topology interacts with dynamics \cite{Gao16,Hens18}, we now understand how the aforementioned factors influence both time- and graph-domain information sampling.

The application domains extend across many engineering, social, and biological complex systems, and we demonstrate our theory on a linear population dynamic (PD) model and a non-linear protein interaction (MAK) model. For recovering the dynamics, our results show that the RMSE drop dramatically by several orders of magnitude when we sample above the optimal sampling rate.

The limitation of our work thus far as been on studying first order one-dimensional Markovian nonlinear dynamics and bandlimited initialization. Many complex systems are a multiplex of different network and dynamics (e.g. multiplexed transport networks \cite{Pagani19}), with dynamics in at least two-dimensions with higher order differential equations (e.g. water distribution networks \cite{Yazdani11} and electricity supply networks \cite{Schafer18}), and have non-Markovian dynamics (e.g. have extended memory of epidemic networks \cite{Scholtes14}). Extending our framework to such network dynamics will be the focus of our future work. \\

\textbf{Author Contributions \& Declarations:} Z.W. developed the theory and conducted the simulations. W.G. developed the idea and suggested the case studies. B.L., Z.W. and W.G. jointly wrote the paper. The authors have no conflict of interests.


\bibliographystyle{IEEEtran}
\bibliography{myref}

\begin{thebibliography}{10}
\providecommand{\url}[1]{#1}
\csname url@samestyle\endcsname
\providecommand{\newblock}{\relax}
\providecommand{\bibinfo}[2]{#2}
\providecommand{\BIBentrySTDinterwordspacing}{\spaceskip=0pt\relax}
\providecommand{\BIBentryALTinterwordstretchfactor}{4}
\providecommand{\BIBentryALTinterwordspacing}{\spaceskip=\fontdimen2\font plus
\BIBentryALTinterwordstretchfactor\fontdimen3\font minus
  \fontdimen4\font\relax}
\providecommand{\BIBforeignlanguage}[2]{{%
\expandafter\ifx\csname l@#1\endcsname\relax
\typeout{** WARNING: IEEEtran.bst: No hyphenation pattern has been}%
\typeout{** loaded for the language `#1'. Using the pattern for}%
\typeout{** the default language instead.}%
\else
\language=\csname l@#1\endcsname
\fi
#2}}
\providecommand{\BIBdecl}{\relax}
\BIBdecl

\bibitem{Gao16}
J.~Gao, B.~Barzel, and A.~Barabasi, ``{Universal resilience patterns in complex
  networks},'' \emph{Nature}, vol. 530, 2016.

\bibitem{Krishnagopal17}
S.~Krishnagopal, J.~Lehnert, W.~Poel, A.~Zakharova, and E.~Scholl,
  ``{Synchronization patterns: from network motifs to hierarchical networks},''
  \emph{Phil. Trans. of the Royal Society A}, vol. 375, 2017.

\bibitem{Barzel13}
B.~Barzel and A.~Barabasi, ``{Universality of Network Dynamics},'' \emph{Nature
  Physics}, vol.~9, 2013.

\bibitem{Wilson08}
A.~Wilson, ``{Boltzmann, Lotka and Volterra and spatial structural evolution:
  an integrated methodology for some dynamical systems},'' \emph{Journal of the
  Royal Society Interface}, vol.~5, 2008.

\bibitem{8105893}
S.~{Dhamal}, R.~D. {Vallam}, and Y.~{Narahari}, ``Modeling spread of
  preferences in social networks for sampling-based preference aggregation,''
  \emph{IEEE Transactions on Network Science and Engineering}, vol.~6, no.~1,
  pp. 46--59, Jan 2019.

\bibitem{Bardoscia17}
M.~Bardoscia, S.~Battiston, F.~Caccioli, and G.~Calderelli, ``{Pathways towards
  instability in financial networks},'' \emph{Nature Communications}, vol.~8,
  2017.

\bibitem{Schafer18}
B.~Schafer, D.~Witthaut, M.~Timme, and V.~Latora, ``{Dynamically induced
  cascading failures in power grids},'' \emph{Nature Communications}, vol.~9,
  2018.

\bibitem{Lu16}
X.~Lu, C.~Gray, L.~Brown, M.~Ledger, A.~Milner, R.~Mondragon, G.~Woodward, and
  A.~Ma, ``Drought rewires the core of food webs,'' \emph{Nature Climate
  Change}, 2016.

\bibitem{Hasegawa18PhyE}
Y.~Hasegawa, ``{Thermodynamics of collective enhancement of precision},''
  \emph{Physical Review E}, vol.~98, 2018.

\bibitem{Scholtes14}
I.~Scholtes, N.~Wider, R.~Pfitzner, A.~Garas, C.~Tessone, and F.~Schweitzer,
  ``{Causality-driven slow-down and speed-up of diffusion in non-Markovian
  temporal networks},'' \emph{Nature Communications}, vol.~5, 2014.

\bibitem{Ellinas17}
C.~Ellinas, N.~Allan, and A.~Johansson, ``{Dynamics of organizational culture:
  Individual beliefs vs. social conformity},'' \emph{PLOS ONE}, vol.~12, 2017.

\bibitem{Duarte12}
M.~Duarte, G.~Shen, A.~Ortega, and R.~Baraniuk, ``{Signal compression in
  wireless sensor networks},'' \emph{Phil. Trans. of the Royal Society A}, vol.
  370, 2012.

\bibitem{Hens18}
C.~Hens, U.~Harush, S.~Haber, R.~Cohen, and B.~Barzel, ``{Spatiotemporal signal
  propagation in complex networks},'' \emph{Nature Physics}, vol.~15, 2019.

\bibitem{chung1996spectral}
F.~R. Chung, ``Spectral graph theory,'' \emph{CBMS regional conference series
  in mathematics}, no.~92, 1996.

\bibitem{p2008}
I.~Pesenson, ``{Sampling in Paley-Wiener spaces on combinatorial graphs},''
  \emph{Transactions of the American Mathematical Society}, vol. 360, no.~10,
  pp. 5603--5627, 2008.

\bibitem{Sandryhaila14}
A.~Sandryhaila and J.~Moura, ``{Discrete signal processing on graphs: frequency
  analysis},'' \emph{IEEE Transactions on Signal Processing}, vol.~62, 2014.

\bibitem{Sandryhaila14M}
------, ``{Big Data Analysis with Signal Processing on Graphs: Representation
  and processing of massive data sets with irregular structure},'' \emph{IEEE
  Signal Processing Magazine}, vol.~31, 2014.

\bibitem{anis2014towards}
A.~Anis, A.~Gadde, and A.~Ortega, ``Towards a sampling theorem for signals on
  arbitrary graphs.'' in \emph{ICASSP}, 2014, pp. 3864--3868.

\bibitem{Chen15}
S.~Chen, R.~Varma, A.~Sandryhaila, and J.~Kovacevic, ``{Discrete Signal
  Processing on Graphs: Sampling Theory},'' \emph{IEEE Transactions on Signal
  Processing}, vol.~63, 2015.

\bibitem{wang2015generalized}
X.~Wang, J.~Chen, and Y.~Gu, ``Generalized graph signal sampling and
  reconstruction,'' in \emph{Signal and Information Processing (GlobalSIP),
  2015 IEEE Global Conference on}.\hskip 1em plus 0.5em minus 0.4em\relax IEEE,
  2015, pp. 567--571.

\bibitem{anis2016efficient}
A.~Anis, A.~Gadde, and A.~Ortega, ``Efficient sampling set selection for
  bandlimited graph signals using graph spectral proxies,'' \emph{IEEE
  Transactions on Signal Processing}, vol.~64, no.~14, pp. 3775--3789, 2016.

\bibitem{Chen16}
S.~Chen, R.~Varma, A.~Sandryhaila, and J.~Kovacevic, ``{Signal Recovery on
  Graphs: Fundamental Limits of Sampling Strategies},'' \emph{IEEE Transactions
  on Signal and Information Processing over Networks}, vol.~4, 2016.

\bibitem{wang2018optimal}
F.~Wang, Y.~Wang, and G.~Cheung, ``{A Optimal Sampling and Robust
  Reconstruction for Graph Signals via Truncated Neumann Series},'' \emph{arXiv
  preprint arXiv:1803.03353}, 2018.

\bibitem{chamon2018greedy}
L.~F. Chamon and A.~Ribeiro, ``Greedy sampling of graph signals,'' \emph{IEEE
  Trans. Signal Process.}, vol.~66, no.~1, pp. 34--47, 2018.

\bibitem{ortega2018graph}
A.~Ortega, P.~Frossard, J.~Kova{\v{c}}evi{\'c}, J.~M. Moura, and
  P.~Vandergheynst, ``Graph signal processing: Overview, challenges, and
  applications,'' \emph{Proceedings of the IEEE}, vol. 106, no.~5, pp.
  808--828, 2018.

\bibitem{8115204}
F.~{Grassi}, A.~{Loukas}, N.~{Perraudin}, and B.~{Ricaud}, ``A time-vertex
  signal processing framework: Scalable processing and meaningful
  representations for time-series on graphs,'' \emph{IEEE Transactions on
  Signal Processing}, vol.~66, no.~3, pp. 817--829, Feb 2018.

\bibitem{Sidropoulos12}
N.~Sidiropoulos and A.~Kyrillidis, ``{Multi-way Compressed Sensing for Sparse
  Low-Rank Tensors},'' \emph{IEEE Signal Processing Letters}, vol.~19, 2012.

\bibitem{Ding17}
X.~Ding, W.~Chen, and I.~Wassell, ``{Joint Sensing Matrix and Sparsifying
  Dictionary Optimization for Tensor Compressive Sensing},'' \emph{IEEE
  Transactions on Signal Processing}, vol.~65, 2017.

\bibitem{lyapunov1992general}
A.~M. Lyapunov, ``The general problem of the stability of motion,''
  \emph{International journal of control}, vol.~55, no.~3, pp. 531--534, 1992.

\bibitem{Pagani19}
A.~Pagani, G.~Mosquera, A.~Alturki, S.~Johnson, S.~Jarvis, A.~Wilson, W.~Guo,
  and L.~Varga, ``{Resilience or Robustness: Identifying Topological
  Vulnerabilities in Rail Networks,},'' \emph{Royal Society Open Science},
  vol.~6, 2019.

\bibitem{Yazdani11}
A.~Yazdani and P.~Jeffrey, ``{Complex network analysis of water distribution
  systems},'' \emph{Chaos}, vol.~21, 2011.

\end{thebibliography}

\end{document}